# Hertz-to-terahertz dielectric response of nanoconfined water molecules

M.Belyanchikov[1], M.Savinov[2], P.Bednyakov[2], Z.Bedran[1], V.Thomas[3], V.Torgashev[4], A.Prokhorov[1,5], A.Loidl[6], P.Lunkenheimer[6], E.Zhukova[1], E.Uykur[7], M.Dressel[1,7], B.Gorshunov[1]

[1] Moscow Institute of Physics and Technology, Dolgoprudny, Moscow Region, 141700 Russia
[2] Institute of Physics AS CR, Na Slovance 2, 18221 Praha 8, Czech Republic
[3] Institute of Geology and Mineralogy, RAS, 630090 Novosibirsk, Russia
[4] Faculty of Physics, Southern Federal University, 344090 Rostov-on-Don, Russia
[5] Prokhorov General Physics Institute of the Russian Academy of Sciences, Moscow, 119991, Russia
[6] Experimental Physics V, Center for Electronic Correlations and Magnetism, University of Augsburg, Universitätsstr. 2, 86135 Augsburg, Germany
[7] 1.Physikalisches Institut, Universität Stuttgart, 70569 Stuttgart, Germany

*Abstract*—Broad-band dielectric spectroscopy, heat capacity measurements and molecular dynamic simulations are applied to study excitations of interacting electric dipoles spatially arranged in a network with an inter-dipole distance of 5-10 Å. The dipoles with magnitude of 1.85 Debye are represented by single $H_2O$ molecules located in voids (0.5 nm size) formed by ions of the crystal lattice of cordierite. We discover emergence of nontrivial disordered paraelectric phase of dipolar system with signs of phase transition below 3 K.

## I. INTRODUCTION

Recently, considerable attention has been given to the properties of interacting electric dipole systems. The electrostatic coupling among the dipoles makes such systems qualitatively different from their *magnetic* counterparts. Since interacting spins have been studied during last decades, significant progress has been achieved in understanding the underlying physics. In *electric* dipole systems, the interplay of quantum tunneling, fluctuations and frustration provides with the possibility to realize exotic phases, like quantum electric dipole liquids and glasses [1], lead to quantum critical phenomena and phase transitions [2]. Understanding the nature of the corresponding phases and their possible relations with magnetic counterparts is of great fundamental and technological interest, but is presently still at its infancy.

An ideal playground for corresponding studies is provided by dielectrics whose crystal lattice contains voids filled with polar molecules that weakly interact with surrounding ions but "feel" each other *via* dipole-dipole interaction. Spectroscopic studies of the gemstone beryl with 0.5 nm sized pores hosting $H_2O$ molecules (each carrying a dipole moment of 1.85 Debye) allowed to discover incipient ferroelectricity within the $H_2O$ dipoles together with a rich set of single-particle excitations at terahertz-infrared frequencies [3-5]. It was suggested that the ferroelectric phase transition is suppressed by quantum tunneling of $H_2O$ molecules within the 1 meV deep localizing hexagonal potential. Here, we study how the properties of the $H_2O$ molecular network can be modified by changing the shape of the potential that holds molecules within the nanopores. Similarly to beryl, orthorhombic cordierite includes channels elongated along the c-axis that contain voids, hosting single $H_2O$ quasi-free molecule. However, in cordierite, the potential is twofold and asymmetric with shape is represented by two approximately

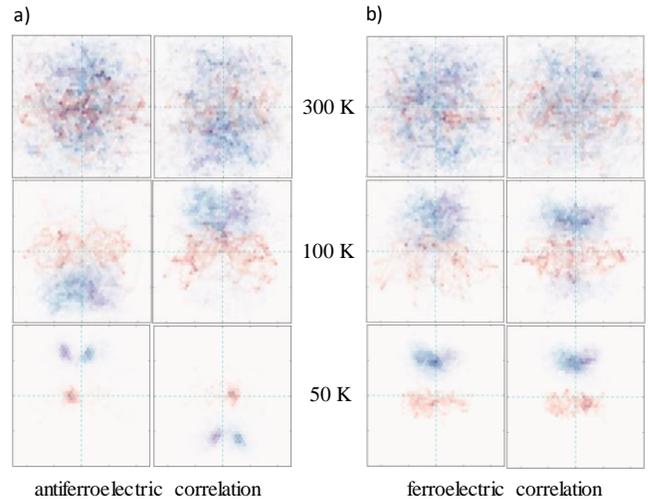

**Fig. 1.** Molecular dynamic DFT simulations of water molecules positions within a cordierite nanopores averaged over 15 ps. a) Two water molecules in nanopores aligned along the c axis are subject to *antiferroelectric* correlations upon cooling, b) Two water molecules in nanopores aligned in the *ab* plane are subject to *ferroelectric* correlations upon cooling. Red color – position of oxygen, blue color – positions of protons.

10 meV deep minima felt by the molecule when it rotates $2\pi$ around c-axis.

## II. EXPERIMENTAL AND COMPUTATIONAL DETAILS

The natural cordierite crystal from India (the detailed location is unknown) were characterized by X-ray scattering and microprobe analysis and cut into 100 – 200 μm slices needed for the polarization-dependent measurements. With radio-frequency (RF), microwave (μW) and terahertz (THz) spectroscopy, we measure polarization-dependent (radiation $E$-field parallel to *a*, *b* and *c* axes) spectra of the complex dielectric permittivity $\varepsilon^*(\nu)=\varepsilon'(\nu)+i\varepsilon''(\nu)$ of hydrous cordierite in the range $\nu$=1 Hz – 3 THz at temperatures 0.3 K – 300 K. At radiofrequencies, we used a Novocontrol Alpha AN High Performance Frequency Analyzer, an Andeen-Hagerling 2500A capacitance bridge and a coaxial reflectometric technique employing an impedance analyzer (Keysight 4991B). For terahertz measurements, a commercial time-domain TeraView 3000 spectrometer was used. The dielectric experiments were complemented by measurements of the heat capacity in the relaxation method employing a PPMS system (Quantum Design). In all experiments, measurements on dehydrated samples allowed us to extract the characteristics determined exclusively by water molecules. To get insight into microscopic background of experimental results we performed molecular dynamic simulations in the framework of density functional theory (DFT) implemented in VASP software package. The 15 ps simulations of water molecules in cordierite were done for three different water filling of unit cell and for six temperature points in the range 10 K – 300 K.

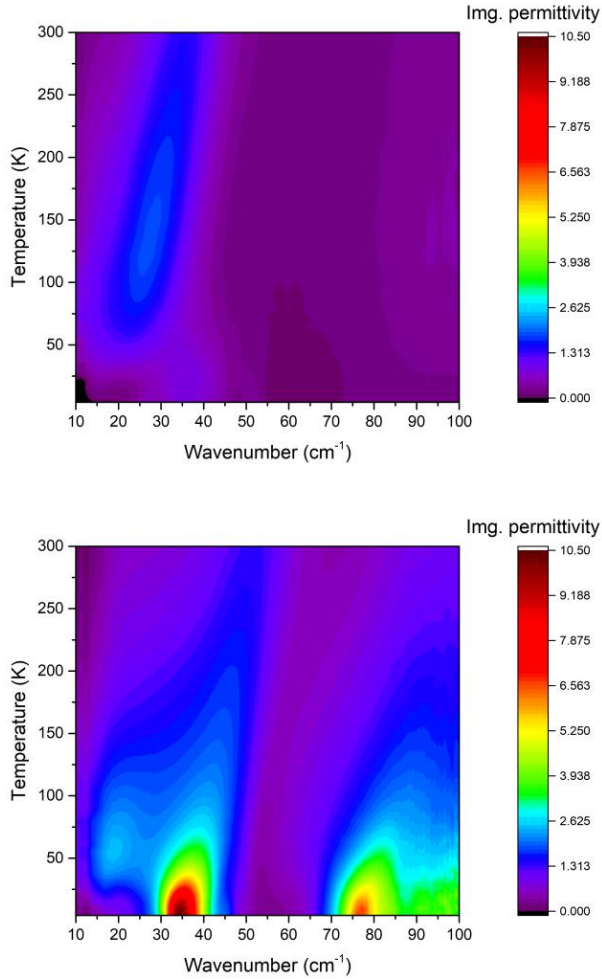

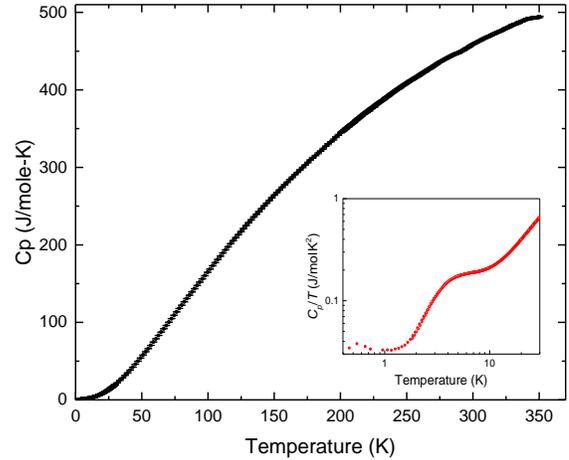

**Fig. 3.** Temperature dependences of specific heat of cordierite crystal with water, $C_p$ graph does not show any sign of a phase transition. The inset shows the temperature dependences of $C_p/T$ with a weak anomaly around 4 K.

**Fig. 2.** 2d map showing temperature evolution of the THz modes (imaginary permittivity) for polarizations $E\|a$ (upper panel) and $E\|b$ (lower panel).

### III. RESULTS AND DISCUSSION

The analysis of the DFT simulations shows that water molecule experience two-well effective potential coming from the crystal lattice environment that leads to double degenerate ground state with a 10 meV energy barrier and appearance of a relaxation peak in microwaves and radio-frequency spectral ranges which will be discussed below. At the same time, when dipole-dipole interaction between water molecules is present, pairs of water molecules start to reveal a correlated manner of their motion during cooling (see Fig. 1). As a consequence of nanopore geometrical arrangement, below $T\approx100$ K the dipole pairs located along and perpendicular to the $c$ axis are subject to antiferroelectric and ferroelectric interactions, respectively. The presence of such interactions leads to appearance of local minima in the potential energy landscape experienced by separate water dipole. In the THz frequency region, several polarization-dependent absorption modes are discovered (see Fig. 2) with nontrivial temperature evolution that is associated with progressive travelling of the dipolar system over the series of local energy minima towards ground state. Despite the presence of ferroelectricaly and antiferroelectricaly correlated dipoles, heat capacity measurement does not show any thermodynamic features in 3 K - 300 K temperature range that could be related to appearing of long-range order (Fig. 3). Hence, down to 3 K, water system in cordierite can be considered as a network of disordered correlated dipole pairs with nontrivial local constraint of locally conserved polarization without stabilized long-range order. Such a state of dipole system can be called a "dipolar ice", based on a similarity of a behavior of ambient hexagonal water ice. There, in the ordered oxygen lattice, the protons have a highly degenerate ground state and there exists specific type of disorder governed by so-called "ice rules" constraints down to helium temperature, or in other words, there exists a zero-point entropy [6].

The system of rotating (around the $c$-axis) dipoles within highly anharmonical potential demonstrates in radio-frequency range typical for paraelectric Curie-Weiss temperature dependence. For the polarization $E\|b$, the complex permittivity spectra are fully determined by soft mode dynamics detected at THz frequencies and exhibit basically no dispersion at all temperatures with saturation of absolute value which is typical for incipient paraelectric. The situation is dramatically different for $E\|a$ polarization, where at lower temperatures, dielectric permittivity shows pronouncedly dispersive behavior due to presence of strongly temperature-dependent overdamped excitation (Fig. 4). The peak frequency of the excitation reveals Arrhenius temperature behavior with two activation energies: 11.2 meV at 7 K<$T$<10 K and 4.6 meV at 4 K<$T$<7 K. At lower temperatures, $T < 3$ K, the behavior of the excitation changes and its frequency *hardens* upon cooling down to 0.3 K. Taking into account the potential energy landscape found from DFT-MD simulations, the absence of the relaxational excitation in a similar water dipole system in beryl crystal [4] and the detected in cordierite anomaly in the water subsystem specific heat (Fig. 3), the detected behavior is interpreted as being caused by a relaxational dynamic of correlated water dipole system within a twofold 10 meV external potential and a minor potential structure at the bottom of potential wells. Below 3 K, a smeared order-disorder transition is happening among dipole system into a state that results in water dipoles tilted out of *bc* pane.

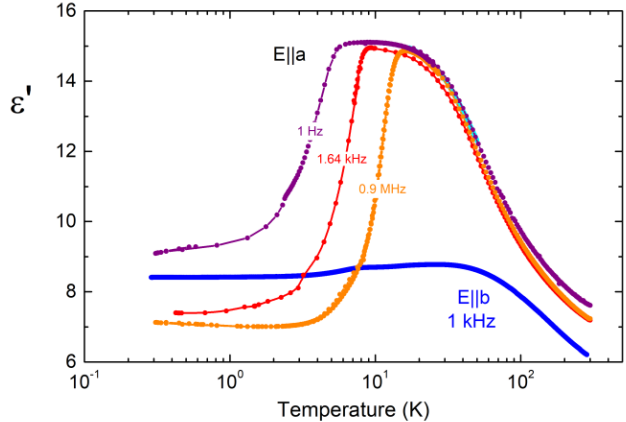

**Fig. 4.** Temperature dependences of real dielectric permittivity of nanoconfined water in mineral cordierite crystal measured at various frequencies for polarizations $E\|a$ and $E\|b$.

## IV. Summary

We have studied the behavior of a system of $H_2O$ electric dipoles that are arranged in a 3D network by localizing them within nanopores formed by ions of cordierite crystal. DFT simulations reveal presence of two types of correlations in macroscopically disordered system of dipoles. Despite the absence of long-range order, even at the lowest temperatures of 0.3 K, the behavior of the dipolar lattice is highly nontrivial and reveals rich sets of polarization-dependent excitations. At temperatures below 3 K, signs of a smeared order-disorder transition are detected.

## V. Acknowledgement

Authors acknowledge support from RFBR 18-32-00286 (data analysis), 18-32-20186 (THz experiments), DFG via DR228/61-1 and Center of Integrated Quantum Science and Technology IQ$^{ST}$ Stuttgart/Ulm (RF- µW experiments).